\documentclass[aps,prl,groupedaddress, twocolumn]{revtex4-1}
\usepackage{graphicx}
\usepackage{dcolumn}
\usepackage{float}
\usepackage{verbatim}
\usepackage{xcolor}
\usepackage{amsmath}
\usepackage{xcolor, colortbl}

\begin{document}


\title{Direct Testing against Experiment of a Fundamental Ultrashort Pulse Laser Damage Simulation Technique with Utility for the Modeling of Nanostructure Formation}


\author{A. M. Russell, K. Kafka, D. W. Schumacher and E. Chowdhury}
\affiliation{}


\date{\today}

\begin{abstract}
We have developed the first laser damage simulation algorithm capable of determining crater and surface modification morphology from microscopic physics. Rapid progress in the field of high intensity ultrafast lasers has enabled its utility in a myriad of applications. Simulation plays an important role in this research by allowing for closer analysis of the physical mechanisms involved, but current techniques struggle to meet both the spatial scope or resolution requirements for modeling such dynamics, typically specializing in one or the other. Consequently, it is difficult to extract the physical form of the laser induced surface modification, hampering direct comparison of simulation to experimental results. Our algorithm offers a compromise to existing simulation techniques and enables the production of a complete density profile in addition to the simulation of intermediate dynamics. We use this capability to directly test our simulation against experimentally produced copper craters. Additionally, we show how our algorithm can be used to model the formation of surface roughness and nanoparticles. 
\end{abstract}

\pacs{}

\maketitle

The development of high intensity ultrashort pulse lasers has enabled the reproducible creation of microscopic surface structures that can usefully alter the electromagnetic and inertial response of materials. As such, they are useful in lithography, machining, development of hydrophobic surfaces, enhancement of optical absorption, and a variety of other applications \cite{short_pulse_lithography, short_pulse_machining, hydrophobic, high_absorption_gratings}. Although the results of such laser target interactions are reproducible, the underlying physics is intrinsically stochastic, making a purely theoretical approach to the underlying physics intractable. Hence, a variety of simulation techniques has been developed for the purpose of predicting experiments and exploring the underlying mechanisms. While modeling interparticle interactions at an atomic scale via molecular dynamics\cite{MD_silica, MD_metal} would be ideal, computational constraints make simulating the entirety of the affected material area via this method difficult. Alternatively, the two temperature model \cite{TTM_ablation} (TTM) is capable of simulating energy deposition and transport on a mesoscopic scale but is incapable of directly determining the physical effect of a laser pulse interaction, instead relying on the critical density to determine ablation properties. This inability to fully simulate the experiment often leads to the prediction of laser damage by experimental data extrapolation, which gives no insight into the physical phenomena involved. We offer a compromise\cite{original_crater_formation} between these two approaches that uses the particle in cell (PIC) method to model the interaction of a material with an ultrashort pulse and the subsequent ablation of matter away from the surface over the entire extent of the laser-affected area. This culminates in a surface modification/damage density profile that can be directly compared to experiment. 

The PIC method is an extremely prevalent simulation technique in the field of plasma physics that operates by directly integrating Maxwell's equations and the Lorentz force law to determine the dynamics of a statistical sampling of particles. It excels at determining the behavior of electromagnetic phenomena over mesoscopic scales, but cannot directly be used for the simulation of a material as it simulates only electric monopoles and thus cannot model the attractive forces that bind materials together. Nevertheless, the infrastructure that it provides and its ability to methodically downsample the resolution and number of particles make it an excellent infrastructure for the modeling of large scale matter transport.

While the kinematics of heated material is a highly complex problem in its own right, for a full analysis of laser induced damage the laser-particle and particle-particle interactions occurring prior to ablation must be investigated in detail as well so that the state of the material is accurately described once the atoms and/or ions in the material attain sufficient energy to mobilize. For the relatively weak leading edge of a damage-inducing pulse, optical properties such as skin depth and absorption are experimentally easy to measure, but the higher energy dynamics such as filling of additional energetic bands, nonthermal behavior, and ballistic particles induced by the remainder of the pulse rapidly complicate an analytic evaluation of these quantities. However, it is possible to characterize these dynamics in a more abstract manner with the temperature and density dependent collision rates, both between ion and electrons ($\nu_{ei}$) and between the electrons themselves  ($\nu_{ee}$). Provided that the laser is sufficiently short, the state of the material immediately after laser impact is a heated population of electrons localized near the surface and within the skin depth and a relatively unperturbed population of ions. Afterward, the phonon-electron interactions, whose main role during the laser pulse was to randomize the motion of electrons, act as a means for temperature equilibration between these two species while the heated electron population continues to diffuse throughout the material. As the phonon energy reaches the atomic disassociation energy, material at the surface begins to mobilize, leading to material flow and ablation.

The problem of modeling laser damage from beginning to end is dramatically simplified by the identification of three different time scales in this process, each of which is modeled separately in our algorithm: laser target interaction (femtosecond), thermalization (picosecond), and ablation (nanosecond). The first of these requires the least adaption as PIC is already equipped to simulate a gaussian beam and contains the requisite microscopic physics for producing reflection. However, PIC's limited spatial resolution neglects the interparticle collisions that lead to velocity randomization and determine the simulated material's absorption, directly affecting the amount and shape of the energy deposition. This physics may be reinserted into PIC by means of a collision algorithm specific to the simulated material. For this work we used the modified binary collision algorithm developed by Russell et al. \cite{Collision_Algorithm} that allows for accurate modeling of absorption in addition to non-thermal effects. With the inclusion of appropriate collision rates and a sufficiently short pulse, an accurate electron heating pattern is produced while the ion temperature remains essentially unmodified. 

Because the electron-ion temperature equilibration happens on a time scale far longer than the laser heating, the simulation of the thermalization process is relegated to the two temperature model, which determines the diffusion and thermalization behavior of the heating profiles for the electrons and ions. When the ions and electrons in the material reach thermal equilibrium, the motion of the former is simulated using the pair potential algorithm. In this final stage we neglect charge separation and model the material as a collection of neutral particles that interact via a pair potential originating from the electron mediated forces between ions. Since applying the pair potential between each pair of particles is too computationally expensive, we instead only apply it between each particle and virtual nearest neighbors, the positions of which are consistent with the density and density gradient. Specifically, we define the virtual nearest neighbor distance $\bar{r}$ via $\bar{r} = n^{-1/3}$ where $n$ is the number density of the particles, effectively approximating a simple cubic lattice. Assuming smoothness of $n$ and $\nabla n$, we may approximate $\bar{r}$ around a particle at location $x_{0}$ as $\bar{r}\left(\mathbf{x}\right) \approx \bar{r}\left(\mathbf{x}_{0}\right) + \nabla\bar{r}\cdot \left(\mathbf{x} - \mathbf{x}_{0}\right)$. Furthermore, we define the nearest neighbor distance $\bar{r}_{\hat{\alpha}}$ in a particular direction $\hat{\alpha}$ as $\bar{r}\left(\mathbf{x}_{0} + \frac{\bar{r}_{\hat{\alpha}}}{2}\hat{\alpha}\right) = \bar{r}_{\hat{\alpha}}$, which indicates that the virtual interparticle distance is equal to the virtual nearest neighbor distance midway between the virtual nearest neighbor and the central particle. From this we find
\begin{equation}
\bar{r}_{\alpha} = \frac{\bar{r}_{0}}{1 - \nabla\bar{r}\cdot\hat{\alpha}}.
\end{equation}
Therefore, the density and density gradient determine virtual nearest neighbor distances to which an appropriate pair potential may be applied. We choose to utilize the Lennard Jones 6-12 potential having the form
\begin{equation}
\label{LJ}
U_{LJ} = D_{e}\left[\left(\frac{r_{eq}}{r}\right)^{12} - 2\left(\frac{r_{eq}}{r}\right)^6\right].
\end{equation}
 The resultant forces determine the trajectories of the neutral atoms and lead to large scale material flow, eventually forming a density profile that can be directly compared to experimental lineouts. 

Despite the copious amount of data on the threshold fluence of ultrashort pulses on a variety of metals \cite{machining_ablation, short_pulse_metal_ablation_1, short_pulse_copper_ablation, short_pulse_metal_ablation_2}, data consisting of actual experimental density profiles is scarce and is spread over a wide variety of fluences, repetition rates and number of pulses. Approximating ablation depth from threshold fluence can be done \cite{threshold_to_ablation_1, threshold_to_ablation_2,threshold_to_ablation_3}, but multiple techniques exist, each of which involves calculations with differing assumptions. Moreover, threshold fluences can differ by as much as an order of magnitude between experiments with negligibly different laser parameters\cite{Axente, short_pulse_metal_ablation_1}, making any crater profile derived from them inherently questionable. To avoid such inconsistencies, we chose to perform our comparison using newly produced copper craters, both for the sake of consistency and because we desired the specifications for our target material, including target preparation, its cut direction, and its single or multi-crystalline nature to be consistent with the simulated material. Furthermore, we wished to tune the spatial and temporal extent of the laser impact so as to reduce computation time. For these reasons, new craters were produced on polished copper cut with a $\langle\!\!$ 100 $\!\!\rangle$ orientation. Single crystalline copper was used to negate the influence of boundary domains on structural integrity and response to the laser. 

\begin{figure}
	\includegraphics[width=1\linewidth]{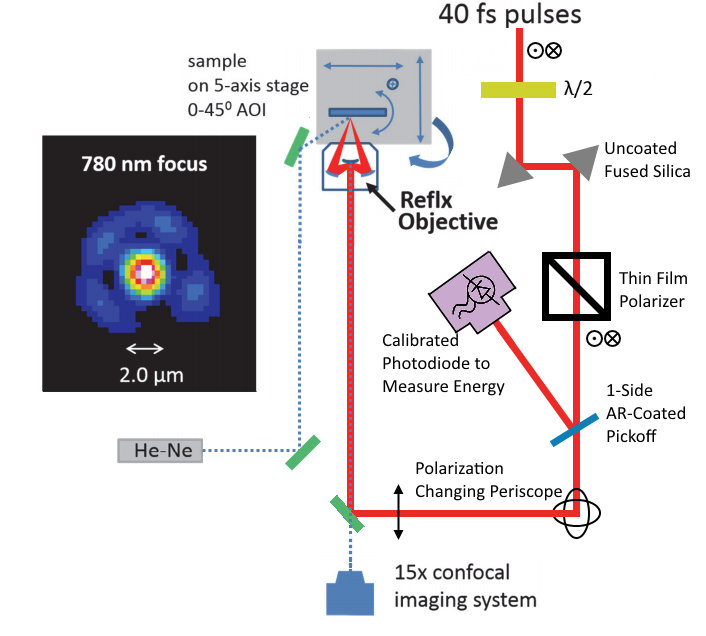}
	\caption{Schematics of experimental setup. Inset shows the laser focal spot generated by the ReflX objective.}
	\label{fig:expsetup}
\end{figure}

\begin{figure*}
	\includegraphics[width=1\linewidth]{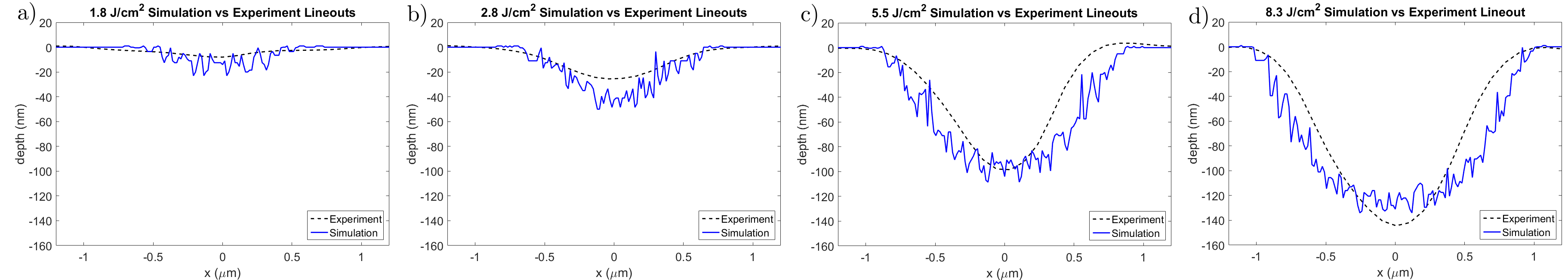}
	\caption{The comparison of experimental lineouts against simulation (a-d). The experimental lineouts are taken along the component of laser propagation parallel to the surface and are shown in blue. The simulation density profile was determined by finding the contour at half solid density and is shown with a dotted black line}
	\label{fig:lineouts}
\end{figure*}

Experiments were performed with pulses from a home-built, Ti:Sapphire, liquid-nitrogen-cooled, regenerative amplifier operating at 500 Hz that produced a 780 nm central wavelength pulse with $>2$ mJ energy compressed to a full width half max (FWHM) of 40 fs. Single pulses were extracted from this train using an external Pockels cell and high-contrast polarizer, and were then sent into the setup shown in Fig. \ref{fig:expsetup}. Energy was first reduced by taking a pickoff reflection from two uncoated fused silica prisms and was then more finely controlled using \textit{an achromatic waveplate and polarizer.} A thin, 1-side AR-coated fused silica window diverted a portion of the pulse to a calibrated photodiode, which allowed in-situ measurement of single-pulse energy. The remaining beam was then focused onto the copper target by a 15x high-performance, infinity-conjugate, reflective objective (Edmund Optics). The focal spot shown in Fig. \ref{fig:expsetup} was obtained via image relaying onto a camera using a second objective and independently measuring the magnification. This focus is primarily a Gaussian beam of waist diameter 2.0 $\mathrm{\mu m}$, with some satellite artifacts due to the nature of the reflective objective. The energy throughput efficiency was also measured, since for a reflective objective the throughput is a function of the input spatial mode distribution and size. The target was aligned with a 5-axis stage to within $\pm0.5$ $\mathrm{\mu m}$ of best focus and at 15 degrees angle of incidence. A HeNe laser illuminated the target surface, assisting in target alignment and enabling in-situ observation of the interaction region. Considering all of the above calibrations, the fluence uncertainty of this experiment is estimated to be $+5/\!\!-\!10\%$, and still produces a crater small enough to directly benchmark our simulation. Crater morphology was measured with an optical profiler (Wyko NT9100 by Veeco) in phase-shift-interference mode (for sub-wavelength depth characterization) for each of the fluences 1.8, 2.8, 5.5 and 8.3 $\mathrm{J/cm}^{2}$. The depth resolution in practice is typically $\lesssim1$ nm, though the lateral resolution is limited by the 535 nm device wavelength. This has the effect of averaging over any nanoscale features $\ll 535$ nm, but should still remain accurate for determination of the envelope of $1\sim2$ $\mathrm{\mu m}$ diameter craters.

The simulation was performed using the three previously described stages with the PIC code LSP \cite{LSP}. In the first the simulated laser parameters were equivalent to experiment while the simulated copper block was $\mathrm{480 \; \mathrm{n m}}$ deep by $4.8 \;\mathrm{\mu m}$ wide. The resolution was 2.5 nm (longitudinal) $\times$ 10 nm (transversal) with a sampling of 9 ions and 900 electrons per cell. Increasing the particle and cell resolution in each direction by $50 \%$ was found not to change the heating pattern by more than $5 \%$. Appropriate collision rates were inserted into the collision model in order to emulate copper's electromagnetic response. The electron-electron collision rate $\nu_{ee}$ was determined via a cubic spline that interpolates between the well known condensed matter and plasma regime, a technique introduced by  Colombier\cite{colombier} et al. For the electron-ion collision rate $\nu_{ei}$ we use those calculated by Lee-More-Desjarlais\cite{LM_Rates, LMD_Rates}, which finds rates via a modified Drude conductivity that has been adapted to more realistically model solids over a wide range of temperatures and densities.

\begin{figure}
	\centering
	\includegraphics[width=1.0\linewidth]{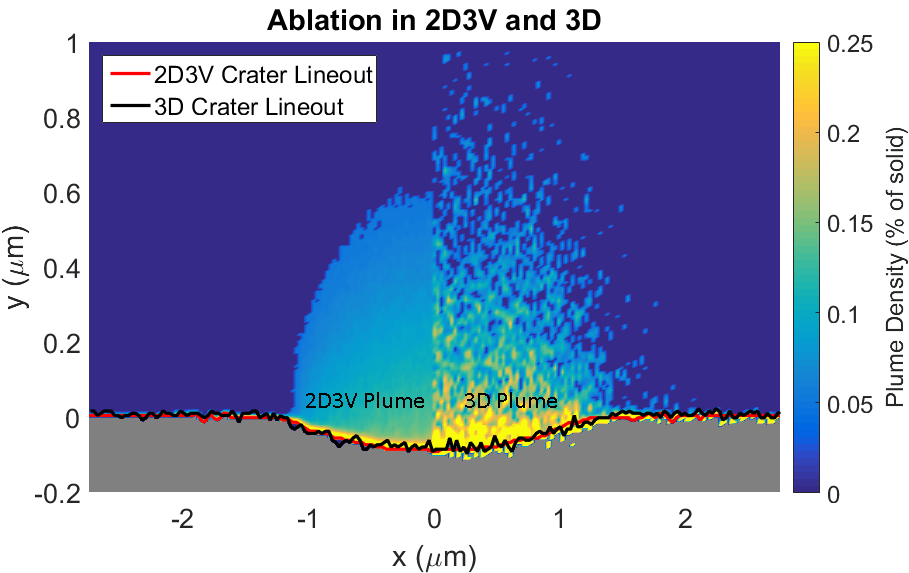}
	\caption{The plume and crater lineout for 2D3V and 3D ablation generated with equivalent heating patterns. The left side of the density plot contains information about the 2D3V ablation and the right side that of the 3D ablation. Zones above half solid density have been colored gray. While the red and black lines indicating the crater lineout describe essentially the same density profile, the ablation plumes differ significantly in their densities near the target. The porous appearance of the 3D plume is an effect of the increased fluctuations due to reduced particle count.}
	\label{fig:Comparison_2D_3D}
\end{figure}

The electron and ion heating profiles were extracted from the first stage immediately after the laser left the simulation and were evolved in time using the TTM. The thermalization routine was run until the electron and ion temperature were within $10\%$ of one another or until $60 ps$ had elapsed, with the latter limit set to avoid restraining ablation. The ion temperature was then extracted and inserted into the last stage of the simulation. The force governing the particle motion was derived from Eq. \ref{LJ}, where the equilibrium interparticle distance was $r_{eq} = .227 \;\mathrm{nm}$ and the disassociation energy was $D_{e} = .3429 \;\mathrm{eV}$. The resolution in this stage was 2.5 nm x 2.5 nm with 900 particles per cell, except for the 1.8 $\mathrm{J/cm^{2}}$ simulation which had a longitudinal resolution of $1.25\;\mathrm{nm}$. Increasing the particle and cell resolution in each direction by $50 \%$ was found to change the density profiles by less than 5 $nm$. After about 1-2 ns the motion of the copper ceased and a density profile was extracted with a cutoff at half solid density. 

In order to reduce computation time, each simulation was run in 2D-3V, meaning that the particles and the grid were restricted to a plane but vectors such as fields, currents, and velocities were 3 dimensional. For the laser target interaction this may be thought of as using a line focus instead of a circular Gaussian beam, where the electric field profile as a function of distance from the center of the pulse is preserved. Heat diffusion in the removed dimension may safely be neglected, as the rate of diffusion perpendicular to the surface of the material versus the rate in directions orthogonal to that scales as $\left(r_{0}/l_{s}\right)^{2}$, where $r_{0}$ is the beam radius and $l_{s}$ is the skin depth. The effect of dimensionality on the third stage was explicitly checked by first calculating a heat pattern generated from a line focused Gaussian beam hitting a target at normal incidence in 2D3V, determining a 3D equivalent by rotating the electron heating pattern around a central axis, and evolving each heating pattern using the pair potential algorithm with the corresponding dimensionality. As shown in Fig. \ref{fig:Comparison_2D_3D}, reduced dimensionality has no significant effect on the crater profile but does change the density profile of the ablated material. Hence, surface morphology may be analyzed in just two spatial dimensions but the analysis of ablated material and of the electromagnetic properties of the surface during the time the plume is present require a fully 3D perspective. 

\begin{figure}
	\includegraphics[width=1\linewidth]{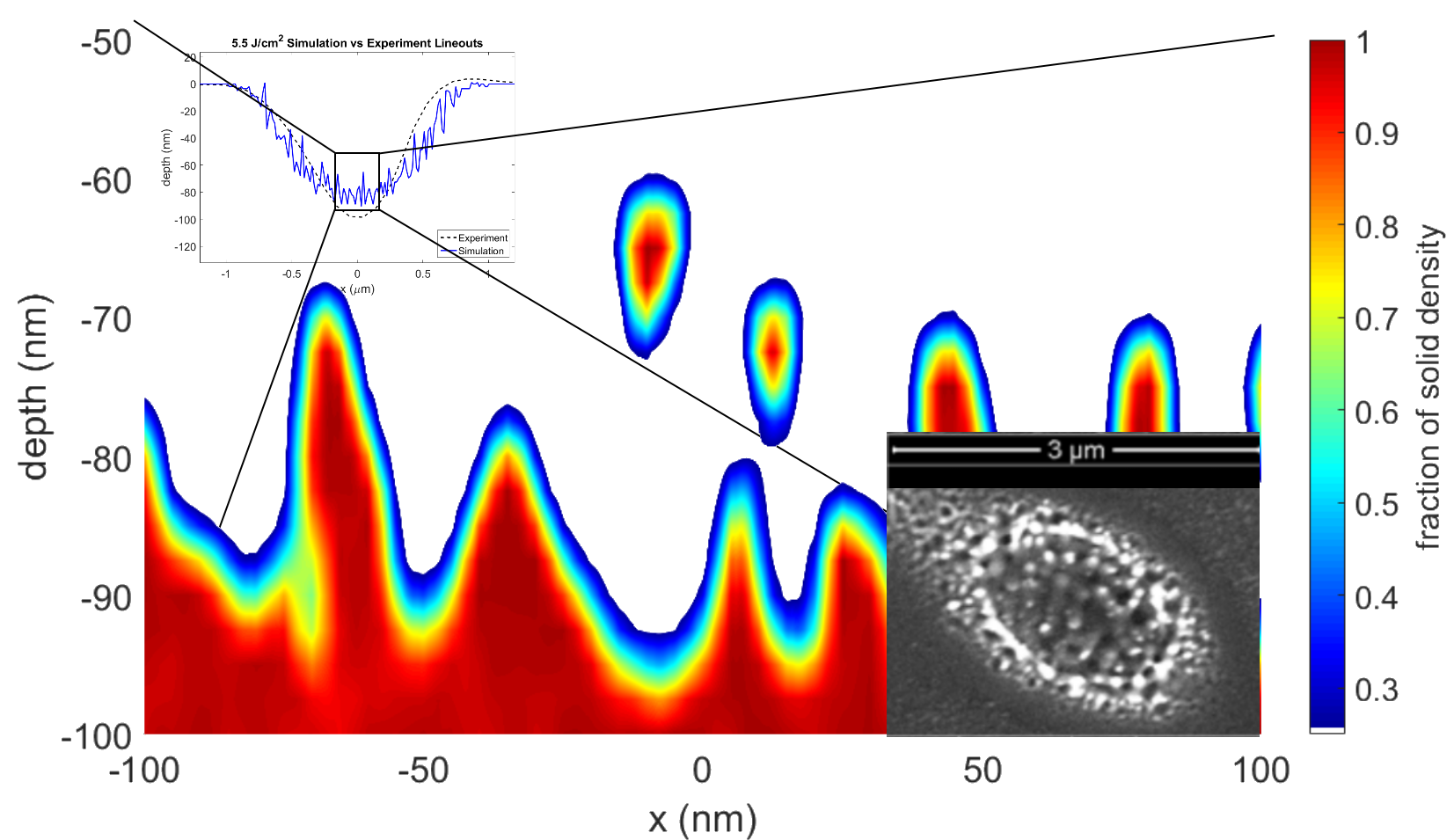}
	\caption{A magnified version of the surface modification seen in the simulation of the 5.5 $\mathrm{J/cm}^{2}$ crater formation. The inset shows an SEM image of the $4.8\; \mathrm{J/cm^{2}}$ experimental crater. Clearly seen in the simulation lineout are both nanostructures and nanoparticles.}
	\label{fig:surface_modification}
\end{figure}

The comparison between experiment and simulation is displayed in Fig. \ref{fig:lineouts}. We see that the general shape and width of the profiles closely match while the depths differ by at most $50 \%$. Much of the error can be attributed to the transition between the second and third stage of the algorithm. Some ablation is likely to occur before the ions and electrons come to thermal equilibrium, which cannot be captured by the TTM, and thermalization is not well captured by the pair potential algorithm. Additionally, there are inaccuracies in the collision rates used and pair potential parameters, the theory for which becomes increasingly tenuous at higher temperatures and densities. The smoothness of the experimental lineout is not real but is instead a consequence of the limited resolution of the Wyko. Fig. \ref{fig:surface_modification} shows the actual texture of the crater. 

While the specifics of nanostructure and microstructure formation are still debated \cite{nanostructure_formation}, it is generally thought to involve the stochastic motion of heated material and, for multipulse systems, the altered electromagnetic properties originating from inchoate surface structures. Our algorithm in particular is ideal for the simulation of these phenomena due to the inclusion of an actual laser and the ability to produce the form of the morphology as an output over the extent of the laser-impacted surface. Surface roughness in particular is visible along the surface of both the simulated and experimental compared craters of Fig.\,\ref{fig:lineouts}. We see in Fig.\,\ref{fig:surface_modification} that the magnified density profile corresponding to the bottom of the crater formed by the 5.5 $\mathrm{J/cm^{2}}$ laser exhibits surface structures on the scale of 10-40 nm in width and 5-20 nm in depth. Shown in the inset is an SEM image recorded from a similar fluence (4.8 $\mathrm{J/cm^{2}}$) which exhibits a similar spatial diameter of individual surface structures of about 100 nm.  Additionally, nanoparticles can be seen several nanometers above the surface in the simulation profile, ranging from 5-10 nm in diameters. The production of nanoparticles via ablation from ultrashort laser pulses has already been studied experimentally \cite{Ausanio} with promising results for the case of nickel, and has the potential to find use in a wide variety of biological and industrial applications \cite{Salata, Santos}.

We have devised an algorithm that models laser damage originating from ultrashort laser pulses by exploiting the infrastructure of the PIC technique commonly used for modeling plasma physics. With it we are able to model the physics of a laser-target interaction, heat diffusion and material ablation from first principles and attain a density profile that can be compared directly to experiment. We perform such a comparison using newly made copper craters and find that the width and depth of the experimental and simulated craters agree to within $50\%$. Moreover, we find that the resolution attainable by our algorithm allows us to simulate the formation of nanoparticles and surface roughness while still capturing mesoscopic features.

\section*{Acknowledgements}

This work was supported by the Ohio Supercomputer Center\cite{OhioSupercomputerCenter1987} and performed under the auspices of the AFOSR grants $\#$ FA9550-16-1-0069 FA9550-12-1-0454.

\bibliography{Benchmarks_Bibliography}

\end{document}